\documentclass[twocolumn,showpacs,preprintnumbers,amsmath,amssymb]{revtex4}
\usepackage{graphicx}
\usepackage{dcolumn}
\usepackage{bm}

\begin{document}

\title{Comment on ``Novel type of phase transition in a system of self-driven particles"}

\author{Michael Z. Q. Chen and Hai-Tao Zhang}
\email{hz254@cam.ac.uk}
\affiliation{Department of Engineering,
University of Cambridge, Cambridge CB2 1PZ, U.K.}
\begin{abstract}
This paper uncovers a pitfall in the phase transition mechanism of Vicsek {\it et~al.} [Phys. Rev. Lett. {\bf 75}, 1226 (1995)] which occurs with fairly high probability and leads to complete breakdown of the model dynamics.
\end{abstract}


\maketitle

In \cite{vi95}, a phase transition model was given in a
system of self-driven particles. Due to the dramatic advances in the emerging and active research of natural and artificial complex networks, the Vicsek model has been drawing more and more attention recently and gaining ascending popularity in the field  \cite{he00}--\cite{al07}. The velocities $\{{\bf v}_i\}$ of the particles were constructed to have an absolute value $v$ and a direction
given by the angle $\theta (t+1)$. This angle was obtained from the
expression 
$\theta (t+1)=\left<\theta\left(t\right)\right>_r+\Delta\theta$,
where
$\left<\theta\left(t\right)\right>_r:=\mbox{arctan}\left[\left<\mbox{sin}\left(\theta\left(t\right)\right)\right>_r/
\left<\mbox{cos}\left(\theta\left(t\right)\right)\right>_r\right]$ 
denotes the average direction of the velocities of particles (including particle $i$) within  radius $r$ surrounding the given particle ($\left<\mbox{sin}\left(\theta\left(t\right)\right)\right>_r$ and
$\left<\mbox{cos}\left(\theta\left(t\right)\right)\right>_r$ being
the average sine and cosine values of the velocities respectively), and $\Delta\theta$ represents a random noise with a uniform probability from the interval $\left[-\eta/2, \eta/2\right]$.

\begin{figure}[htb]
  \centering \leavevmode
     \resizebox{7cm}{!}{\includegraphics[width=11.2cm]{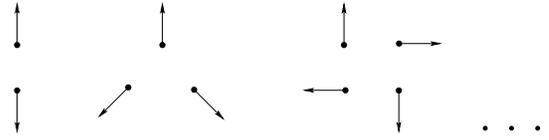}}
 \caption{Cases of direction annihilation.}
 \label{fig:special cases}
 \end{figure}

However, in bacterial colony, flocks of birds, schools of fish and multi-particle systems, it is fairly possible that the effects of all particles counteract completely within a radius of $r$. We define such a phenomenon as {\it direction annihilation}.  Some simple examples are given in Fig.~\ref{fig:special cases}.  It can be easily shown that the possibility of such cases occurring descends with increasing number of particles. 

\begin{figure}[htb]
  \centering \leavevmode
     \resizebox{7cm}{!}{\includegraphics[width=11.2cm]{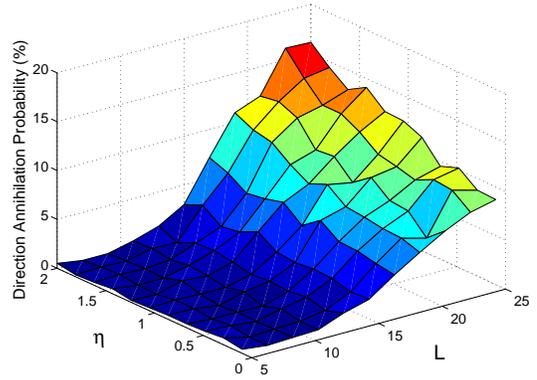}}
 \caption{Direction annihilation probability.}
 \label{fig:direction annihilation probability}
 \end{figure}

Since $0/0$ is not properly defined, the phase transition mechanism in \cite{vi95} breaks down when 
$\label{special case}\left<\mbox{sin}\left(\theta\left(t\right)\right)\right>_r
=\left<\mbox{cos}\left(\theta\left(t\right)\right)\right>_r=0$ (i.e. direction annihilation) within a radius of $r$ around a particular particle. The probability of direction annihilation occurring is statistically simulated over $1000$ runs for each $\eta$ and $L$ and shown in 
Fig.~\ref{fig:direction annihilation probability} with $L$ being the linear size of a square cell. It can be observed that the lower the density, the more likely that direction annihilation can occur, which agrees with the fact that the possibility of direction annihilation occurring descends with increasing number of particles. Moreover, the probability of direction annihilation increases with the intensity enhancement of the noise as the noise tends to break the synchronised direction of particles therefore increase the possibility of direction annihilation. As direction annihilation occurs with a fairly high probability and it potentially jeopardises all the processes and research works employing the Vicsek Model (\cite{he00}--\cite{al07}), it is necessary to overcome this problem in \cite{vi95}. 

We propose three possible transition rules governing the self-driven particle system: keep the original direction, add one most adjacent particle outside the radius $r$ or drop one farthest particle within the radius. The latter two are equivalent to slightly increasing and decreasing the radius respectively. Any of the above three rules is sufficient to overcome the problem in Vicsek's Model.
 
\end{document}